\documentclass[paper]{JHEP3} 

\JHEPspecialurl{http://jhep.sissa.it/JOURNAL/JHEP3.tar.gz}

\usepackage{epsfig,multicol}

\newcommand\fverb{\setbox\pippobox=\hbox\bgroup\verb}
\newcommand\fverbdo{\egroup\medskip\noindent%
            \fbox{\unhbox\pippobox}\ }
\newcommand\fverbit{\egroup\item[\fbox{\unhbox\pippobox}]}
\newcommand{\id}{{1\!\!1}} 
\newcommand {\beq}{\begin{equation}}
\newcommand {\eeq}{\end{equation}}
\newcommand {\beqa}{\begin{eqnarray}}
\newcommand {\eeqa}{\end{eqnarray}}
\newcommand {\n}{\nonumber \\}
\newcommand {\tr}{{\rm tr\,}}
\newcommand {\Tr}{\mbox{Tr\,}}

\newcommand {\ee}{\mbox{e}}

\newbox\pippobox                                                                   %
\title{Supersymmetric lattice gauge theories in
noncommutative geometry}
\title{Supersymmetry on the Noncommutative Lattice}

\author{Jun Nishimura ${}^a$, Soo-Jong Rey ${}^{b,c}$,
Fumihiko Sugino ${}^b$\\
Department of Physics, Nagoya University \\
Furo-cho, Chikusa-ku, Nagoya 464-8602 {\rm JAPAN} ${}^a$\\
School of Physics \& BK-21 Physics Division \\
Seoul National University, Seoul 151-747 {\rm KOREA} ${}^b$\\
School of Natural Sciences, Institute for Advanced Study\\
Einstein Drive, Princeton NJ 08540 {\rm USA} ${}^c$
\\
E-mail: \email{nisimura@eken.phys.nagoya-u.ac.jp,
sjrey,sugino@phya.snu.ac.kr} }

\preprint{SNUST-030102\\{\tt hep-lat/0301025}}  

\abstract{ \\
Built upon the proposal of Kaplan {\sl et.al.} [hep-lat/0206109],
we construct noncommutative lattice gauge theory with manifest
supersymmetry. We show that such theory is naturally implementable
via orbifold conditions generalizing those used by Kaplan {\sl
et.al.} We present the prescription in detail and illustrate it
for noncommutative gauge theories latticized partially in two
dimensions. We point out a deformation freedom in the defining
theory by a complex-parameter, reminiscent of discrete torsion in
string theory. We show that, in the continuum limit, the
supersymmetry is enhanced only at a particular value of the
deformation parameter, determined solely by the size of the
noncommutativity.}

\keywords{Matrix Model, Lattice Gauge Theory, Supersymmetry,
Duality}

\begin{document}


\section{Introduction}
Matrix models have played an important role in string and gauge
theories. They provide a powerful constructive approach to
two-dimensional quantum gravity coupled to various matters, which
can also be viewed as noncritical string theories. This idea has
developed into the recent conjectures that a certain type of
matrix models, called as reduced models, serve as nonperturbative
definitions of superstring/M theories \cite{IKKT,Banks:1996vh}. In
particular there are some evidences that four-dimensional
space-time is {\em dynamically} generated in the matrix model for
type IIB superstring theory \cite{4dfromiib}.

Historically, the reduced model was introduced
as an equivalent description of large $N$ gauge theories
\cite{EK,GO,Gonzalez-Arroyo:1983ac} in the planar limit. It can be
formally obtained by dimensional reduction of SU($\infty$) gauge
theories. Recently reduced models have been reinterpreted as gauge
theories (with gauge group of finite rank) on noncommutative
geometry \cite{CDS,Aoki:1999vr,AMNS1,AMNS2,AMNS3}. In this
interpretation, the space-time coordinates and the color indices
are treated on equal footing as matrix indices. The gauge
invariance results from the SU($\infty$) symmetry of the reduced
model. Making the size of the matrices finite corresponds to
discretizing the space-time into a lattice
\cite{AMNS1,AMNS2,AMNS3}. This approach was crucial for rendering
field theories on noncommutative geometry accessible by Monte
Carlo simulations \cite{sim_nc}. By further imposing appropriate
orbifold conditions, one can obtain a finite noncommutative torus
with an arbitrary noncommutativity parameter \cite{AMNS1}. As a
particular case, one can also obtain commutative space-time. Thus,
any lattice field theory (including Wilson's lattice gauge theory)
can be embedded in a matrix model with a certain orbifold
condition.

Recently, reduced models were shown to be also useful for
constructing lattice theories with manifest supersymmetry
\cite{Kaplan:2002wv}. The construction consists of two steps.
First one considers the mother theory, which is a reduced model
obtained from dimensional reduction of SU($N$) super Yang-Mills
theory {\em in the continuum}. (In fact the case with maximal
supersymmetry corresponds to the matrix models for superstring/M
theories mentioned above.) Then one imposes the orbifold condition
on the mother theory and arrives at a daughter theory, which
inherits part of the supersymmetry from the mother theory. Here,
the orbifold condition plays a crucial role in introducing a
lattice structure to the daughter theory. The idea of respecting
some symmetry on the lattice and avoiding fine-tuning thereof in
obtaining a supersymmetric continuum limit was also discussed in
Refs.\ \cite{lattice_susy}.

In this paper, we consider generalization of Kaplan {\sl et.al.}'s
new approach to lattice supersymmetry and show that noncommutative
super Yang-Mills theories can be naturally constructed by adopting
a generalized orbifold condition (similar to the consideration of
Ref.\ \cite{AMNS1} in a different context). As an illustration, we
present an explicit construction of a noncommutative U($k$) gauge
theory with maximal supersymmetry, \footnote{It was speculated
that this sort of theories accommodates gravity as well, though
they are defined on a flat (noncommutative) space-time
\cite{Ishibashi:2000hh}.} which is of particular importance due to
its relationship to superstring/M theories \cite{CDS,Aoki:1999vr}.
We also mention a close relationship to discrete torsion studied
in the context of string theory on orbifold.

\section{Emergent Space-Time out of Matrix Orbifolds}
\label{orbifold_st}
We begin with a brief recapitulation concerning emergent
space-time out of matrices via certain orbifold projection
conditions. As a toy model, consider a matrix model of $(N \times
N)$ complex matrices $\Phi_i$ ($i=1,\cdots , M$), whose action is
given  by
\beq S = {1 \over g^2} \Tr (\Phi_1 \cdots \Phi_M ).
\label{matrix_action} \eeq
On the matrices $\Phi_i$, we impose `orbifold conditions' of the
following sort:
\beq \Phi_i = {\omega_L}^{r_{i,a}} \, \Omega_a^\dagger \Phi_i \,
\Omega_a \ . \label{orbifold_cond} \eeq
Here, $\omega_L$ is a phase-factor
\beqa \omega_L := \ee^{ 2 \pi i \over L} \qquad {\rm obeying}
\qquad {\omega_L}^L = 1, \nonumber \eeqa
and $\Omega_a$ ($a = 1, \cdots , d $) are $N \times N$ unitary
matrices. Details of the emergent space-time, including
(non)commutativity, turn out to depend on specific choices of
these matrices. For now, we choose them to be
\beqa \Omega_1 &=& U_L \otimes \id_L \otimes \cdots \otimes \id_L
\otimes  \id_k \n \Omega_2 &=& \id_L \otimes U_L \otimes \cdots
\otimes \id_L \otimes  \id_k \n ~ &\vdots& ~ \n \Omega_d &=&
\underbrace{\id_L \otimes \id_L \otimes \cdots \otimes U_L }_{d}
\otimes  \id_k  \ . \label{orbifold_mat} \eeqa
We assume that $N$, the matrix size, can be factorized as
\beq N = k \cdot L^d \  \nonumber \eeq
for some integer $k$. In the definition Eq.(\ref{orbifold_mat}),
$\id_L$ and $\id_k$ stand for a unit matrix of the specified
sizes, and $U_L$ stands for the `clock matrix' of size $p=L$:
\beq U_p=\pmatrix{1& & & & \cr &\omega_p& & & \cr& &{\omega_p}^2&
& \cr& & &\ddots& \cr & & & &{\omega_p}^{p-1}\cr} \ .  \label{clock}
\eeq
For later convenience, we also introduce the `shift matrix' by
\beq V_p=\pmatrix{0&1& & & &0\cr &0&1& & & \cr& &\ddots&\ddots& &
\cr& & &\ddots& &1\cr 1& & & & &0\cr} \ . \label{shift} \eeq
The pair of matrices Eqs.(\ref{clock},\ref{shift}) satisfies the
well-known 't Hooft-Weyl algebra:
\beqa U_p V_p=\omega^{-1}_p \,V_pU_p \qquad {\rm where} \qquad
\omega_p := e^{ 2 \pi i \over p} \ .  \label{weylalgebra} \eeqa

For $\Phi_i$'s obeying the orbifold condition
Eq.(\ref{orbifold_cond}), the set of ``charge vector'' $({\bf
r}_i)_a := r_{i,a}$, assigned uniquely for each matrix $\Phi_i$,
ought to satisfy the condition:
\beqa \sum_{i=1}^M {\bf r}_i = 0 \ , \label{0sum} \eeqa
else the action Eq.(\ref{matrix_action}) would vanish trivially.
We refer Eq.(\ref{0sum}) as `charge neutrality condition', and, in
what follows, we will assume that it is always satisfied. In
constructing super-symmetric gauge theories, a convenient choice
of ${\bf r}_i$ would be to take a suitable linear combination and
re-scaling of the charges associated with the R-symmetry
\cite{Kaplan:2002wv}.

To solve Eq.(\ref{orbifold_cond}), we find it convenient to
introduce $(N \times N)$ unitary matrices $D_a$:
\beqa D_1 &=& V_L^\dagger \otimes \id_L \otimes \cdots \otimes
\id_L \otimes \id_k\n D_2 &=& \id_L \otimes V_L^\dagger \otimes
\cdots \otimes \id_L \otimes  \id_k\n ~ &\vdots& ~ \n D_d &=&
\underbrace{\id_L \otimes \id_L\otimes \cdots \otimes
V_L^\dagger}_d \otimes  \id_k\ , \nonumber \eeqa
which obey a different set of orbifold conditions
\beqa D_b = {\omega_L}^{\delta_{a,b}} \, \Omega_a^\dagger D_b
\Omega_a \qquad (a,b = 1, \cdots, d) \ . \nonumber \eeqa
Then, by virtue of the `t Hooft-Weyl algebra
Eq.(\ref{weylalgebra}), we find a particular solution to the
orbifold condition Eq.(\ref{orbifold_cond}) as
\beqa \Phi^{(0)}_i = \prod_{a=1}^d {D_a}^{r_{i,a}} \ . \nonumber
\eeqa
We then decompose the matrices $\Phi_i$'s around the particular
solution:
\beqa \Phi_i = \widetilde{\Phi}_i \Phi^{(0)}_i \ , \label{docomp}
\eeqa
and find that the shifted matrices $\widetilde{\Phi}_i$'s obey
`homogeneous' orbifold condition:
\beq \widetilde{\Phi}_i  = \Omega^\dagger_a \widetilde{\Phi}_i
\Omega_a \ . \label{hom_orbifold_cond} \eeq

As the conditions Eq.(\ref{hom_orbifold_cond}) are linear
equations, all we need is to construct a complete set of basis of
the (finite-dimensional) solution space. For this purpose, we
shall introduce $(L^d \times L^d)$ unitary matrices \footnote{Note
that, in the present case, $Z_a\otimes \id_k$'s turn out identical
to $\Omega_a$'s, but they are distinct matrices in general cases.
We will encounter such a case later when constructing
noncommutative space-time.}, denoted as $Z_a$'s:
\beqa Z_1 &=& U_L \otimes \id_L \otimes \cdots \otimes \id_L  \n
Z_2 &=& \id_L \otimes U_L \otimes \cdots \otimes \id_L \n ~
&\vdots& ~ \n Z_d &=& \underbrace{\id_L \otimes \id_L \otimes
\cdots \otimes U_L}_d \ . \nonumber \eeqa
As defined so, $(Z_a)^L = 1$ for all $a=1, \cdots, d$.

Evidently, any matrices of the form $Z_a \otimes M$, where $M$ is
an arbitrary $(k \times k)$ matrix, are solutions to the
homogeneous orbifold condition Eq.(\ref{hom_orbifold_cond}). For
the special case of $k=1$, the complete basis of the solution
space is spanned by
\beqa J({\bf p}) = \prod_{a=1}^d (Z_a)^{p_a} \qquad \qquad {\rm
where} \qquad {\bf p} := (p_1, \cdots, p_d)\ , \nonumber \eeqa
and $p_a$'s take values in $0,1,\cdots , (L-1)$. A dual basis,
which turns out more convenient, is obtainable via the Fourier
transformation as
\beqa \Delta ({\bf n}) := \sum_{\bf p} J({\bf p}) \,
{\omega_L}^{{\bf p}\cdot {\bf n} } \qquad {\rm where} \qquad {\bf
n} :=( n_1, \cdots, n_d) \ . \nonumber \eeqa
Here again, $n_a$'s take values in $0,1,\cdots , (L-1)$. For
arbitrary $k > 1$, a general solution to the homogeneous orbifold
condition Eq.(\ref{hom_orbifold_cond}) is always expressible as
\beqa \widetilde{\Phi}_i = \sum_{\bf n} \Delta({\bf n}) \otimes
\varphi_i({\bf n}) \ , \label{matrix_field} \eeqa
where $\varphi_i({\bf n})$'s denote $(k\times k)$ matrix-valued
functions of the $d$-dimensional lattice vector ${\bf n}$.
Substituting this and Eq.(\ref{docomp}) into the mother theory
action Eq.(\ref{matrix_action}) and utilizing the identity
\beqa D_a \left[\Delta ({\bf n}) \otimes \varphi_i({\bf n})
\right] D_a^\dagger = \Delta ({\bf n}-\hat{a}) \otimes
\varphi_i({\bf n})\ , \nonumber \eeqa
we finally obtain an action of the daughter theory:
\beq S= \sum_{\bf n} \tr \Bigl[ \varphi_1({\bf n}) \varphi_2({\bf
n}+ {\bf r}_{1} ) \varphi_3({\bf n}+ {\bf r}_1 + {\bf r}_2) \cdots
\varphi_M ({\bf n}+ {\bf r}_1 + \cdots + {\bf r}_{M-1} ) \Bigr] \
. \label{lattice_action} \eeq
Here, the symbol `$\tr$' is to denote the trace over the $(k
\times k)$ matrices, as contrasted to the symbol `$\Tr$' in
Eq.(\ref{matrix_action}) denoting the trace over the $(N \times
N)$ matrices. Accordingly, the matrix degrees of freedom are
reduced by the ratio, $N/k = L^d$, which is precisely the volume
of the emergent space-time. Hence, the zero-dimensional matrix
model Eq.(\ref{matrix_action}), along with the orbifold condition
Eq.(\ref{orbifold_cond}) is the same as a (matrix) field theory on
the $d$-dimensional lattice,
where the matrix-valued functions $\varphi_i({\bf n})$ brought up
in Eq.(\ref{matrix_field}) are interpreted as (matrix) fields on
the emergent lattice.

At the outset, we have assumed that the charge vectors obey the
neutrality condition Eq.(\ref{0sum}). What is the reason behind
the condition? What would happen if the condition is not met?
Notice that, prior to imposing the orbifold condition
Eq.(\ref{orbifold_cond}), the mother theory
Eq.(\ref{matrix_action}) is actually invariant under U($N$)
rotation:
\beq \Phi_i \quad \longmapsto \quad G \, \Phi_i \, G^\dag \qquad
{\rm where} \qquad G \in {\rm U}(N). \label{SU_N_tr} \eeq
This invariance is compatible with the orbifold condition
Eq.(\ref{orbifold_cond}) for $\Phi_i$ if and only if the rotation
matrix $G$ satisfies the homogeneous orbifold condition
Eq.(\ref{hom_orbifold_cond}) as well. In that case, in complete
analogy with Eq.(\ref{matrix_field}), we can write the orbifold
projected rotation matrix $G$ as
\beqa G = \sum_{\bf n} \Delta({\bf n}) \otimes g({\bf n}) \ .
\nonumber \eeqa
Inserting this back to Eq.(\ref{SU_N_tr}), we readily find that
the U($N$) rotation is reduced to a U($k$) {\sl gauge}
transformation
\beqa \varphi_i({\bf n}) \quad \longmapsto \quad g({\bf n}) \,
\varphi_i({\bf n}) \, g^\dag({\bf n}+ {\bf r}_i) \ .
\label{gauge_tr} \eeqa
Now, Eq.(\ref{gauge_tr}) indicates that, if the lattice field
$\varphi_i({\bf n})$ carries no charge (${\bf r}_i = 0$), it may
be considered naturally as a variable residing on the lattice site
${\bf n}$, while those with charge (${\bf r}_i \ne 0$) may be
considered as a field residing on the lattice link connecting site
${\bf n}$ and site $({\bf n}+ {\bf r}_i)$. The latter fields are
precisely the counterpart of link variables in the standard
lattice gauge theory. We then learn that the daughter theory
Eq.(\ref{lattice_action}) is invariant under the emergent gauge
transformation Eq.(\ref{gauge_tr}) if and only if the lattice
links close up, viz.
\beqa {\bf n} = {\bf n} + {\bf r}_1 + \cdots + {\bf r}_{M-1} \ .
\nonumber \eeqa
We thus learn that the charge neutrality condition Eq.(\ref{0sum})
is precisely what ensures the emergent daughter theory to possess
the local gauge invariance Eq.(\ref{gauge_tr}).

In the above construction, the emergent space-time was commutative
because all the $Z_a$'s and hence all the $\Delta({\bf n})$'s
commute one another. This originates from the specific choice of
the $\Omega_a$ matrices as in Eq.(\ref{orbifold_mat})
\footnote{Our choice of $\Omega_a$'s in Eq.(\ref{orbifold_mat}) is
the same as the one taken by Kaplan {\sl et.al.}
\cite{Kaplan:2002wv}.}. If we adopt different choices of the
$\Omega_a$ matrices, we may be able to render the emergent
space-time noncommutative. What specific choices are required for
the (non)commutative space-time is then an interesting question,
and we obtain the answer in detail in the next sections.
\section{Super Yang-Mills Theory On The Commutative Tori}
In this section, we will present a prescription of constructing
theories with manifest supersymmetry on a commutative tori. We
will do so in a set-up generalizable to noncommutative case. For
concreteness, we will consider a $(2+1)$-dimensional super
Yang-Mills theory with sixteen supercharges and gauge group
U($N$). We will first describe the commutative case in some detail
in this section, and generalize it to the noncommutative case in
the next section.

\subsection{The Mother Theory}
For technical convenience, we prescribe the mother theory as
follows. Consider the $(3+1)$-dimensional ${\cal N}=4$ U($N$)
super Yang-Mills theory, which may be expressed in terms of the
manifest ${\cal N}=1$ superfield notation (See, e.g., Ref.\
\cite{WB} for conventions and notations). Dimensionally reducing
it down to $(0+1)$ dimension, we arrive at the mother theory with
the action $S = \int {\rm d} t \, L$, where the Lagrangian $L$ is
given as
\beqa L & = & L_g + L_{\Phi} + L_W, \n L_g & = & \frac{1}{16g^2}\,
\Tr \, \left. \Big[W^{\alpha}W_{\alpha}\Big]\right|_{\theta\theta}
+ \mbox{h.c.}, \n
L_{\Phi} & = & \frac{1}{g^2}\sum_{a = 1}^3 \Tr \left.
\Big[\overline{\Phi}_a e^{\cal V}\Phi_a e^{-{\cal V}}\Big]
\right|_{\theta\theta\bar{\theta}\bar{\theta}}, \n
L_W & = & \left. \frac{\sqrt{2}}{g^2} \, \Tr \,
\Big[\Phi_1\Phi_2\Phi_3 - \Phi_1\Phi_3\Phi_2
\Big]\right|_{\theta\theta} + \mbox{h.c.} \ . \label{mother} \eeqa
${\cal V}$ is a vector superfield and $\Phi_j$ are chiral
superfields. $W^{\alpha}$ is a superfield containing the gauge
field strength, which is made from ${\cal V}$. (Before dimensional
reduction, $L_g$ represents (3+1)-dimensional ${\cal N}=1$ pure
super Yang-Mills theory, while $L_{\Phi}$ contains kinetic terms
and gauge interaction terms of six scalar fields. The
$(\mbox{commutator})^2$-interactions among the six scalar fields
come out of $L_W$ after integrating out auxiliary fields. See
Appendix A.)

The mother theory Eq.(\ref{mother}) inherits the ${\rm SU}(4)$
R-symmetry of the (3+1)-dimensional ${\cal N}=4$ super Yang-Mills
theory. Under the R-symmetry group rotation, six components (real
and imaginary parts) of $\Phi_1, \Phi_2, \Phi_3$ transform as the
antisymmetric representation ${\bf 6}$, whereas the fermions
transform chirally in the fundamental representation ${\bf 4}$.
Combined with the three extra scalar multiplets arising from
dimensional reduction from $(3+1)$- to $(0+1)$-dimensions, the
R-symmetry group of the mother theory is promoted to $G^{\rm
mother}_R ={\rm Spin}$(9). We shall be interested in the quotient
of the R-symmetry group after imposing a given choice of orbifold
conditions.

\subsection{The Daughter Theory}
As we are going to construct two spatial dimensions out of
matrices, we shall be considering the $d=2$ case of the orbifold
condition Eq.(\ref{orbifold_cond}). Therefore, the size $N$ of
each matrix in the mother theory is taken as $N=k \cdot L^2$. The
orbifold group in Eq.(\ref{orbifold_cond}) is $G_O = {\bf Z}_L
\otimes {\bf Z}_L$. R-charge vectors of the mother theory fields
are denoted as ${\bf r}_s (s = v, 1, 2, 3)$, and are assigned with
integer-valued components as in the following table. This amounts
to picking up three Cartan subgroup
$SO(2) \times SO(2) \times SO(2)$ of
$G^{\rm mother}_R={\rm Spin}$(9), and use
two independent combinations of them for the phase-rotation in
the orbifold conditions Eq.(\ref{orbifold_cond}). As in the
previous section, the orbifold group $G_O$ lets a two-dimensional
lattice emerge in the end. Hereafter, we adopt a unified notation
$\Phi_v \equiv {\cal V}$.
\TABLE[pos]{
\\
\\
\begin{tabular}{c|c}
\hline \hline
$\Phi_s$     & $ {\bf r}_s = (r_{1}, r_{2})_s$ \\
\hline
${\cal V}$   & $(\,\,0\,\,, \,\,0\,\,)$   \\
$\Phi_1$     & $(+2, \,0\,\,)$   \\
$\Phi_2$     & $(-1, +1)$  \\
$\Phi_3$     & $(-1, -1)$ \\
\hline \hline
\end{tabular}
\label{table}
}
Imposing the orbifold conditions Eq.(\ref{orbifold_cond}), we
obtain a daughter theory given by
\beqa L_g & = & \frac{1}{16g^2}\sum_{\bf n} \tr
\left.\Big[W^{\alpha}({\bf n})W_{\alpha}({\bf n})
\Big]\right|_{\theta\theta} + \mbox{h.c.}, \n
L_{\Phi} & = & \frac{1}{g^2} \sum_{\bf n} \tr \Big[
\overline{\Phi}_1({\bf n}) e^{{\cal V}({\bf n})} \Phi_1({\bf n})
e^{-{\cal V}({\bf n} + 2\hat{x})} \n
 & & \hspace{1.5cm}+ \overline{\Phi}_2({\bf n})
e^{{\cal V}({\bf n})}
\Phi_2({\bf n}) e^{-{\cal V}({\bf n} -\hat{x}+\hat{y})}  \n
 & & \hspace{1.5cm} \left.\left.
+\overline{\Phi}_3({\bf n}) e^{{\cal V}({\bf n})}
\Phi_3({\bf n}) e^{-{\cal V}({\bf n} -\hat{x}- \hat{y})}\right]
\right|_{\theta\theta\bar{\theta}\bar{\theta}}, \n
L_W & = & \frac{\sqrt{2}}{g^2}\sum_{\bf n} \tr \Big[ \Phi_1({\bf
n})\Phi_2({\bf n} + 2\hat{x}) \Phi_3({\bf n} +\hat{x} +\hat{y}) \n
 & & \hspace{1.4cm} \left. -\Phi_1({\bf n})\Phi_3({\bf n} + 2\hat{x})
\Phi_2({\bf n} +\hat{x} -\hat{y})\Big]\right|_{\theta\theta} \n
&+& {\rm h.c.}. \label{daughter1} \eeqa
Here, $\hat{x}$ and $\hat{y}$ denote unit vectors along the
emergent $x$- and $y$- directions. After the orbifold projection,
the number of super-symmetry is reduced to four from sixteen the
mother theory Eq.(\ref{mother}) originally retained.

A parameter defining the daughter theory is $L$. If $L$ is even,
the theory comprises of two decoupled daughter theories defined on
even and odd lattice sites, respectively, and thus we end up with
two copies of the same theory. Here, we assume that $L$ is odd, in
which case the resulting theory is defined on the entire lattice.
Using the periodicity($n_x \sim n_x + L$), we can interpret that
$n_x$ takes the even values
\beqa n_x = 0,2,\cdots, L-1, L+1, L+3, \cdots, 2L-2 \label{n1}
\eeqa
for sites of $n_y$ even, and that $n_x$ takes the odd values
\beqa n_x = 1,3,\cdots, L-2, L, L+2, \cdots, 2L-1 \label{n2} \eeqa
for sites of $n_y$ odd. As a consequence of the interpretation,
all the interactions on the lattice in Eq.(\ref{daughter1})
connect only nearest neighbor lattice sites.

The boundary conditions of the lattice fields turn out somewhat
nontrivial:
\beq \Phi_s({\bf n} +2L\hat{x}) = \Phi_s({\bf n} + L\hat{x} +
L\hat{y}) = \Phi_s({\bf n}). \label{BC} \eeq
In the $y$-direction, the simple periodic boundary condition
$\Phi_s({\bf n} +L\hat{y}) = \Phi_s({\bf n})$ does not hold. Note
that ${\bf n} + L\hat{y}$ does not belong to the lattice sites if
${\bf n}$ is a lattice site. The boundary condition Eq.(\ref{BC})
indicates that the two-dimensional lattice space is a torus
obtained by identifying opposite edges of a parallelogram
connecting the sites $(0,0)$, $(2L, 0)$, $(L, L)$, $(3L, L)$.

To render the parallelogram isotropic, we introduce re-scaled
lattice spacings for $x$- and $y$-directions as $\epsilon_x =
\frac12\epsilon$ and $\epsilon_y = \frac{\sqrt{3}}{2}\epsilon$.
The two-dimensional lattice now consists of equilateral triangles.
Denote site coordinates as ${\bf x} = (x, y) = (n_x\epsilon_x,
n_y\epsilon_y)$.
Denote also $\ell\equiv L\epsilon$. We also denote the lattice
fields as
\beq \Phi_s({\bf n})=
\Phi_s(n_x, n_y)\equiv \Phi_s(x,y) = \Phi_s({\bf x}). \eeq The
boundary conditions Eq.(\ref{BC}) then become
\beq \Phi_s({\bf x} +
\ell\hat{x}) = \Phi_s\left({\bf x} + \frac12 \ell\hat{x} +
\frac{\sqrt{3}}{2} \ell\hat{y}\right) = \Phi_s({\bf x}). \eeq
Finally, the daughter theory Lagrangian becomes
\beqa L_g & = & \frac{1}{16g^2}\sum_{\bf x} \tr
\left.\Big[W^{\alpha}({\bf x})W_{\alpha}({\bf x})
\Big]\right|_{\theta\theta} + \mbox{h.c.}, \n
L_{\Phi} & = & \frac{1}{g^2} \sum_{\bf x} \tr \Big[
\overline{\Phi}_1({\bf x}) e^{{\cal V}({\bf x})} \Phi_1({\bf x})
e^{-{\cal V}({\bf x} + {\bf m}_1\epsilon)}  +
\overline{\Phi}_2({\bf x}) e^{{\cal V}({\bf x})} \Phi_2({\bf x})
e^{-{\cal V}({\bf x} + {\bf m}_2\epsilon)} \n
 & & \hspace{1.4cm} \left.
+\overline{\Phi}_3({\bf x}) e^{{\cal V}({\bf x})} \Phi_3({\bf x})
e^{-{\cal V}({\bf x} +{\bf m}_3\epsilon)}\Big]
\right|_{\theta\theta\bar{\theta}\bar{\theta}}, \n
L_W & = & \frac{\sqrt{2}}{g^2}\sum_{\bf x} \tr \Big[ \Phi_1({\bf
x})\Phi_2({\bf x} + {\bf m}_1\epsilon) \Phi_3({\bf x} -{\bf
m}_3\epsilon) \n
 & & \hspace{1.5cm} \left.-\Phi_1({\bf x})\Phi_3({\bf x} + {\bf m}_1\epsilon)
\Phi_2({\bf x} -{\bf m}_2\epsilon)\Big]\right|_{\theta\theta} +
\mbox{h.c.}, \label{daughter2} \eeqa
where
\beqa {\bf m}_1  & = & \hat{x},  \n {\bf m}_2  & = &
-\frac12\hat{x} + \frac{\sqrt{3}}{2}\hat{y}, \n {\bf m}_3  & = &
-\frac12\hat{x} - \frac{\sqrt{3}}{2}\hat{y}. \nonumber
\eeqa
Note that the rescaled R-charge vectors satisfy charge neutrality
condition:
\beqa {\bf m}_1 + {\bf m}_2 + {\bf m}_3 = 0 \ , \nonumber \eeqa
and the three charge vectors form an equilateral triangle.

The R-symmetry group of the daughter lattice theory is $G_R^{\rm
daughter} = [{\rm U}(1)]^3 \times {\rm Spin}(3)$, which is a
subgroup of $G_R^{\rm mother}=$ Spin(9) surviving after the
orbifolding. $[{\rm U}(1)]^3$ is individual phase-rotation
symmetry of $\Phi_j$ $(j=1,2,3)$ accompanied by a suitable
rotation of $\theta$, compensating the total phase. Spin(3)
rotates the three scalars (and their superpartners) in ${\cal V}$.
The diagonal part of the $[{\rm U}(1)]^3$:
\beqa \Phi_j \rightarrow e^{i2\beta/3}\Phi_j \qquad {\rm with}
\qquad \theta \rightarrow e^{-i\beta}\theta \label{diagonalU(1)}
\eeqa
fits with the interpretation that the emergent space-time after
the orbifolding is a two-dimensional lattice --- each lattice
direction breaks the supersymmetry by one-half, so one-quarter of
the sixteen supercharges (transforming under Spin(7) as ${\bf
8_s}$) would be preserved.
\subsection{Continuum and Infinite Volume Limit}
We now consider the continuum and infinite volume limits of the
theory Eq.(\ref{daughter2}). In terms of component fields,
Eq.(\ref{daughter2}) is expressed as
\beqa L_g & = & \frac{1}{g^2} \sum_{\bf x} \tr \Big[
i\bar{\lambda}({\bf x}){\cal D}_0 \lambda({\bf x}) -
\bar{\lambda}({\bf x})\sigma^j[a_j({\bf x}), \lambda({\bf x})] +
\frac12 D({\bf x})^2 + \frac12 \left({\cal D}_0 a_j({\bf x})
\right)^2  \n
 & & \hspace{1.5cm}
+ \frac14 [a_j({\bf x}), a_l({\bf x})]^2 \Big], \n
L_{\Phi} & =& \frac{1}{g^2} \sum_{\bf x} \tr \Big[ \left|F_j({\bf
x})\right|^2 + \left|{\cal D}_0B_j({\bf x}) \right|^2 +
i\overline{\psi}_j({\bf x}){\cal D}_0 \psi_j({\bf x}) \n
 & & - \Big|a_l({\bf x})B_j({\bf x})-B_j({\bf x})a_l({\bf x}+
{\bf m}_j\epsilon)\Big|^2 -\overline{\psi}_j({\bf x})\sigma^l
\left\{a_l({\bf x})\psi_j({\bf x}) - \psi_j({\bf x})a_l({\bf x} +
{\bf m}_j\epsilon) {}^{} \right\} \n
 & & + \Big(i\sqrt{2}\overline{B}_j({\bf x})
\left\{\lambda({\bf x})\psi_j({\bf x}) - \psi_j({\bf x})
\lambda({\bf x} + {\bf m}_j\epsilon) \right\} + {\rm h.c.} \Big)
\n
 & & + D({\bf x}) \left\{B_j({\bf x})\overline{B}_j({\bf x}) -
\overline{B}({\bf x}-{\bf m}_j\epsilon)B_j({\bf x} - {\bf
m}_j\epsilon) \right\} \Big], \n
L_W & = & \frac{\sqrt{2}}{g^2}\sum_{\bf x} \tr \Big[ F_1({\bf x})
\left\{ B_2({\bf x}+{\bf m}_1\epsilon)B_3({\bf x} - {\bf
m}_3\epsilon) -B_3({\bf x}+ {\bf m}_1\epsilon)B_2({\bf x} - {\bf
m}_2\epsilon) \right\} \n
 & & \hspace{1.5cm} - B_1({\bf x})
\left\{ \psi_2({\bf x} + {\bf m}_1\epsilon)\psi_3({\bf x} - {\bf
m}_3\epsilon) - \psi_3({\bf x}+{\bf m}_1\epsilon)\, \psi_2({\bf
x}-{\bf m}_2\epsilon) \right\} \n
&& \hskip1.5cm + (1 \rightarrow 2, \,\, 2 \rightarrow 3, \,\, 3
\rightarrow 1) \n && \hskip1.5cm + (1 \rightarrow 3, \,\, 2
\rightarrow 1, \,\, 3 \rightarrow 2) \Big] \n
 & & + \mbox{h.c.},
\eeqa
where $j$, $l$ run over 1,2, and 3,  $a_j$ are real scalar fields
originating from spatial components of the $(3+1)$-dimensional
gauge fields before dimensionally reduced to Eq.(\ref{mother}).

We expand the complex scalar fields $B_j$ around the vacuum
configuration $(f/\sqrt{2}) \id_k$ as
\beqa B_j({\bf x}) = (f /\sqrt{2}) \id_k + B_j'({\bf x}) \qquad
{\rm with} \qquad f = \sqrt{2/3}{\epsilon}^{-1} \ .
\label{shifting} \eeqa
As the expectation value of $B_j$'s is proportional to the unit
matrix, the gauge group U($k$) remains unbroken. Thanks to the
manifest supersymmetry on the lattice, the degrees of freedom are
balanced between bosons and fermions. Thus, it is expected that
the theory is free from the problem of fermion doubling. In fact,
as pointed out in Ref. \cite{Kaplan:2002wv}, the vacuum
expectation value of $B_j$'s induces fermion bilinear terms (out
of the Yukawa coupling terms in $L_\Phi$) which take the form of
the Wilson fermion mass term (with a particular value of the
Wilson coupling parameter, a point further elaborated in
\cite{rozali}).

Continuum limit is taken with \footnote{Then we have to fix two
independent radion modes (constant modes of the U(1) part of
$\frac{1}{\sqrt{3}}h_x+ \frac{1}{\sqrt{6}}h_3$ and $h_y$) as
analyzed by Kaplan {\sl et.al.} \cite{Kaplan:2002wv}.}
\beqa \epsilon \rightarrow 0, \qquad \ell\equiv L\epsilon =
\mbox{fixed}, \qquad g_3^2 \equiv \epsilon_x\epsilon_y g^2 =
\mbox{fixed} \ . \nonumber \eeqa
The volume of the resulting two-dimensional torus is $\ell\times
\frac{\sqrt{3}}{2}\ell$, so, if necessary, an infinite volume
limit is attainable by taking $\ell \rightarrow \infty$, while the
noncommutativity is held fixed \cite{mandalreywadia}.

In order to get the theory in the standard form, it is convenient
to make the field redefinition
\beqa & & B_j'\equiv  \frac{1}{\sqrt{3}}~{\bf m}_j\cdot({\bf h} +
i{\bf v}) + \frac{1}{\sqrt{6}}~(h_3 + ih_4), \hspace{1cm}
\overline{B}_j' \equiv \frac{1}{\sqrt{3}}~{\bf m}_j\cdot({\bf h} -
i{\bf v}) + \frac{1}{\sqrt{6}}~(h_3 - ih_4), \n & & \psi_j \equiv
\sqrt{\frac23}~{\bf m}_j\cdot {\bf \Psi} + \frac{1}{\sqrt{3}}~\xi,
\hspace{3.55cm} \overline{\psi}_j \equiv  \sqrt{\frac23}~{\bf
m}_j\cdot \overline{{\bf \Psi}} + \frac{1}{\sqrt{3}}~\xi,
\nonumber \eeqa
with
\beqa {\bf h} = \left(\begin{array}{c} h_x \\ h_y \end{array}
\right), \hspace{1cm} {\bf v} = \left(\begin{array}{c} v_x \\ v_y
\end{array} \right), \hspace{1cm} {\bf \Psi} =
\left(\begin{array}{c} \psi_x \\ \psi_y \end{array} \right),
\hspace{1cm} \overline{{\bf \Psi}} = \left(\begin{array}{c}
\bar{\psi}_x \\ \bar{\psi}_y \end{array} \right). \nonumber \eeqa
The component fields $v_x$, $v_y$, $h_x$, $h_y$, $h_3$, $h_4$ are
hermitian matrices, and $\psi_x$, $\psi_y$, $\xi$ are complex
2-component spinors.

Finally, we arrive at the following continuum theory in
$(2+1)$-dimensions:
\beqa L_g & = & \frac{1}{g_3^2} \int {\rm d}^2x ~ \tr
\left[i\bar{\lambda}{\cal D}_0\lambda -
\bar{\lambda}\sigma^j[a_j, \lambda] + \frac12 D^2 + \frac12 ({\cal
D}_0 a_j)^2 + \frac14 [a_j, a_l]^2  \right], \n
L_{\Phi} & = & \frac{1}{g_3^2} \int {\rm d}^2x ~ \tr\left[
\left|F_j \right|^2 + \frac12 \left\{
\left(F_{0i}\right)^2 - ({\cal D}_i a_j)^2 \right\} + \frac12
\Big\{ ({\cal D}_0 h_I)^2
 + [a_j, h_I]^2 \Big\} \right. \n
 & &
+ i\overline{\psi}_x{\cal D}_0\psi_x + i\overline{\psi}_y{\cal
D}_0\psi_y + i\overline{\xi}{\cal D}_0\xi
-i\{\psi_i{\cal D}_i\lambda -\overline{\psi}_i{\cal
D}_i\overline{\lambda}\} \n
 & &
-\{\overline{\psi}_i\sigma^j[a_j, \psi_i]
+i\lambda[h_i, \psi_i] +i\overline{\lambda}[h_i,
\overline{\psi}_i]\} \n
 & & \left. \frac{}{}
-\overline{\xi}\sigma^j[a_j, \xi] -\lambda[h_4+ih_3, \xi]
-\overline{\lambda}[h_4-ih_3, \overline{\xi}] +D \left({\cal
D}_xh_x + {\cal D}_yh_y -i[h_3, h_4] \right) \right], \n
L_W & = & \frac{1}{g_3^2}\int d^2x ~~ \tr \Big[
F_1\frac{1}{\sqrt{6}}\{ iF_{xy} + {\cal D}_xh_y-{\cal D}_yh_x +
\sqrt{2}~{\cal D}_y(h_3 + ih_4) +[h_x, h_y] \n
 & & \hspace{3.5cm} + \sqrt{2}~[h_y, h_3 + ih_4] \} \n
 & & +
F_2\frac{1}{\sqrt{6}}\left\{ iF_{xy} + {\cal D}_xh_y -{\cal D}_yh_x -
\sqrt{\frac32}~{\cal D}_x(h_3+ih_4) -
\frac{1}{\sqrt{2}}~{\cal D}_y(h_3 +ih_4) +[h_x, h_y] \right. \n
 & & \hspace{2cm}\left.-\sqrt{\frac32}~[h_x, h_3+ih_4]-
\frac{1}{\sqrt{2}}~[h_y, h_3 + ih_4] \right\} \n
 & & +
F_3\frac{1}{\sqrt{6}}\left\{ iF_{xy} + {\cal D}_xh_y -{\cal D}_yh_x +
\sqrt{\frac32}~{\cal D}_x(h_3+ih_4) -
\frac{1}{\sqrt{2}}~{\cal D}_y(h_3 + ih_4) +[h_x, h_y] \right. \n
 & & \hspace{2cm}\left.+\sqrt{\frac32}~[h_x, h_3+ih_4]-
\frac{1}{\sqrt{2}}~[h_y, h_3 + ih_4] \right\} \n
 & & \left. \frac{}{}
-\xi{\cal D}_x\psi_y + \xi{\cal D}_y\psi_x -\xi[h_x, \psi_y] +
\xi[h_y, \psi_x] + \psi_x[h_3 + ih_4, \psi_y]\right] \n
 & & + \mbox{h.c.},
\label{continuum} \eeqa
where the indices run as $i = x,y$, $I = x,y,3,4$. Also, $j, l =
1,2,3$, as before. $F_{0i}$, $F_{xy}$ are the gauge field strength
$F_{0i} = \partial_0v_i - \partial_i v_0 + i[v_0, v_i]$, $F_{xy} =
\partial_xv_y-\partial_yv_x +i[v_x, v_y]$, and
$a_j$, $h_I$ stand for 7 scalars. After integrating the auxiliary
fields $D$ and $F_j$'s, it is straightforward to see that the
theory is the same as the standard form of (2+1) dimensional
super-Yang-Mills theory.

In the continuum limit, the supersymmetry is enhanced (quadrupled)
and preserves the sixteen supercharges. In Eq.(\ref{continuum}),
the manifest R-symmetry is Spin(7), under which the seven scalars
$a_j, h_I$ transform as the vector representation ${\bf 7}$ and
the fermions $\psi_x, \psi_y, \xi, \lambda$ transform as the
spinor representation ${\bf 8}$ \footnote{Note that, as Kaplan
{\sl et.al.} \cite{Kaplan:2002wv} argued, the four supercharges
manifest on the lattice was sufficient to recover the full sixteen
supercharges in the continuum limit without any fine-tuning.}. An
important point is that, in the continuum limit, we are referring
the R-symmetry Spin(7) to the symmetry involving the {\sl shifted}
scalar fields $B_j'$ in Eq.(\ref{shifting}), whereas the
R-symmetry $[{\rm U}(1)]^3 \times {\rm Spin}(3)$ of the lattice
daughter theory Eq.(\ref{daughter2}) were the one concerning the
$B_j$'s descended from the mother theory. Note that the
expectation value in Eq.(\ref{shift}) breaks the diagonal U(1)
symmetry Eq.(\ref{diagonalU(1)}). At a generic vacuum of the
continuum theory, where the gauge group is broken to $[{\rm
U}(1)]^k$, the photon is equivalent (via duality transformation)
to a scalar field. Combined with the existing seven scalar fields,
the continuum daughter theory would exhibit the continuum limit
R-symmetry Spin(8).
\section{Super Yang-Mills Theory On The Noncommutative Lattice}
We start again with the mother theory Eq.(\ref{mother}), but now
construct a noncommutative version of the lattice super-Yang-Mills
theory by a different choice of the orbifold conditions.
\subsection{The Daughter Theory}
Here, we take $N = k \cdot mq \cdot nq$, where $k,m,n,q$ are
integers, and denote $I:=mnq$. We take orbifold conditions for the
fields as
\beq \Phi_s  = {\omega_{nq}}^{r_{s,a}} \, \Omega_a^\dag \, \Phi_s
\, \Omega_a \qquad \mbox{for all} \qquad s,a \ ,
\label{NC_orbifolding} \eeq
where $\omega_{nq}$ denotes the $(nq)$-th root of the unity. We
now take the orbifold condition matrices $\Omega_{a} \in {\rm
U}(N)$ as~\footnote{These choices were considered first in
\cite{AMNS1} for obtaining noncommutative space-time.}
\beqa \Omega_1 & = & {U_I}^m \otimes {V_q^\dagger}^{p} \otimes
{\id}_k \n
\Omega_2 & = & \, {V_I}^m \otimes \, U_q^{\dagger} \, \otimes
{\id}_k \ . \nonumber \eeqa
In the present case, unlike the mutually commutative ones
Eq.(\ref{orbifold_mat}) that rendered commutative daughter
theories, $\Omega_{a}$'s do not commute each other, but obey the
`t Hooft-Weyl algebra
\beqa \Omega_1\Omega_2 = e^{2\pi i\Theta}\Omega_2\Omega_1 \qquad
{\rm where} \qquad \Theta = {1 \over q} \left(p -
\frac{m}{n}\right) \quad ({\rm mod} \,\, 1). \nonumber \eeqa
The R-charge vectors ${\bf r}_{s}$ in Eq.(\ref{NC_orbifolding})
are assigned the same as in the table in section \ref{table}.

We now introduce $D_{a} \in {\rm U}(N)$ as
\beqa D_1  =  V_I^{\dagger} \otimes \id_q \otimes \id_k \qquad
{\rm and} \qquad
D_2  =  U_I \otimes \id_q \otimes \id_k, \nonumber \eeqa
with the property
\beqa D_{a}\Omega_{b} = {\omega_{nq}}^{-\delta_{ab}} \,
\Omega_{b}D_{a}. \nonumber \eeqa
We then define shifted matrix fields $\widetilde{\Phi}_s$ as
\beqa \Phi_s = \widetilde{\Phi}_s D_1^{r_{s, 1}}D_2^{r_{s, 2}}
\qquad \qquad (s=v, 1,2,3) \nonumber \eeqa
so that $\widetilde{\Phi}_s$'s are subject to {\sl homogeneous}
orbifold conditions:
\beqa \widetilde{\Phi}_s = \Omega_{a} \widetilde{\Phi}_s
\Omega_{a}^{\dagger} \qquad \mbox{for all} \qquad s,a \ .
\label{NC_zero_orbifolding} \eeqa
These orbifold conditions are solvable as follows. Assume that $p$
and $q$ are co-prime, and $r$ and $s$ are integers such that
$rp+sq = 1$. We introduce $Z_{a} \in {\rm U}(mq \cdot nq)$ defined
as
\beqa Z_1 =  {U_I}^n \otimes V_q^\dagger \qquad {\rm and} \qquad
Z_2 = {V_I}^{n} \otimes {U_q^\dagger}^r \  \nonumber \eeqa
with the periodicity:
\beqa {Z_1}^L = {Z_2}^L= \id_{mq \cdot nq}  \qquad {\rm where}
\qquad L := mq \ . \nonumber \eeqa
 The $Z_a$'s have algebraic properties that they commute with orbifold
 condition matrices:
\beqa \left[\Omega_{a}, Z_b \right] = 0 \qquad {\rm for}
\,\,\,{\rm all} \quad a,b \ , \nonumber \eeqa
but $Z_{a}$'s do not commute with each other:
\beqa Z_1Z_2 = e^{- 2\pi i \Theta'}Z_2Z_1 \qquad {\rm where}
\qquad \Theta' = {1 \over q} \left(\frac{n}{m} -  r \right) \quad
(\mbox{mod }1) \ .
\nonumber \eeqa
 We can then construct the complete set of basis for a general solution
 of the homogeneous orbifold condition Eq.(\ref{NC_zero_orbifolding}) as
\beqa J({\bf m}) = e^{-\pi i\Theta'm_1m_2} \,  {Z_2}^{m_2}
{Z_1}^{m_1}  = J^\dagger(-{\bf m}). \label{defJ} \eeqa
 Here, $m_{a}$'s run over $m_{a} = 0,1,\cdots,
L-1$. We shall be imposing periodic boundary conditions $J({\bf
m}+L\hat{a}) = J({\bf m})$ for every $a$-th direction. Due to the
phase-factor $e^{-\pi i\Theta'm_1m_2}$ introduced in
Eq.(\ref{defJ}) for manifest hermiticity, the periodicity requires
that $L\Theta'$ is an even-integer. Taking the case $L$ is
odd\footnote{ In the case $L$ even, there does not exist $\Theta'$
satisfying the periodicity for $(n-mr)$ odd.}, we can then choose
$\Theta'$ as
\beqa \Theta' & = & {1 \over q} \left(\frac{n}{m} -r \right)
\hskip1.4cm (\mbox{mod }2) \qquad \mbox{when $(n-mr)$ is even.} \n
\Theta' & = & {1 \over q} \left( \frac{n}{m} - r \right) +1
\hskip0.8cm (\mbox{mod }2) \qquad \mbox{when $(n-mr)$ is odd.}
\nonumber \eeqa
The dual basis to configuration space is obtained via
\beqa \Delta({\bf n}) = \sum_{m_1, m_2} J({\bf m}) \,
{\omega_L}^{{\bf m} \cdot {\bf n}}, \nonumber \eeqa
where again ${\bf n} =(n_1, n_2)$ spans a two-dimensional lattice,
ranging in each direction over $0, 1, \cdots, L-1$.
As is constructed, $\Delta({\bf n})$ is hermitian and periodic
$\Delta({\bf n}+L\hat{a}) = \Delta({\bf n})$ along each $a$-th
direction.

To proceed further, we note that the dual basis $\Delta({\bf n})$
satisfies the following identities:
\beqa
 & & \frac{1}{L^2} \sum_{\bf n}\Delta({\bf n}) = \id_{mq\cdot nq}, \n
 & & D_{a}\left[\Delta({\bf n})\otimes \varphi({\bf n})\right]
 D_{a}^{\dagger} = \Delta({\bf n} - \hat{a})\otimes \varphi({\bf
n}), \n
 & & \widehat{\tr} \left[\Delta({\bf n}) \right] = mq \cdot nq, \n
 & & \widehat{\tr} \left[\Delta({\bf n})\Delta({\bf n}') \right] =
 \delta_{{\bf n}, {\bf n}'} \, mq\cdot nq \, L^2, \label{ids} \eeqa
where $\varphi({\bf n})$ is arbitrary $k\times k$ matrix.
`$\widehat{\tr}$' denotes the trace for $(mq\cdot nq) \times (mq
\cdot nq)$ matrices.

As in the commutative case, the dual basis $\Delta({\bf n})$ spans
a basis for the subspace complement to the last factor of
$(k\times k)$ matrices, so we can decompose the (shifted) matrix
variables of the mother theory as
\beq
\widetilde{\Phi}_s = \frac{1}{L^2} \sum_{\bf n}
\Delta({\bf n})\otimes\widehat{\Phi}_s({\bf n})\ .
\label{NC_general_solution}
\eeq
Here, $\widehat{\Phi}_s$'s are $(k\times k)$ matrix-valued fields
defined on the emergent two-dimensional lattice labelled by ${\bf
n}$.
Eq.(\ref{NC_general_solution}) then provides a general solution to
the homogeneous orbifold condition Eq.(\ref{NC_zero_orbifolding}),
and hence, after taking into account of shifts through $D_a$'s, to
the orbifold condition Eq.(\ref{NC_orbifolding}).

One last thing we will need to understand concerns how matrix
multiplication in the mother theory is translated in the daughter
theory. To answer this, consider $N\times N$ matrices $f_1$ and
$f_2$ that commute with $\Omega_{a}$'s. As $\Delta({\bf n})$ spans
a complete set of basis, the matrices are Fourier-decomposable as
in Eq.(\ref{NC_general_solution}):
\beqa f_i = \frac{1}{L^2}\sum_{\bf n} \Delta({\bf n})\otimes
\widehat{f}_i({\bf n}) \qquad i = 1, 2 \ . \nonumber \eeqa
Product of two matrices is again a matrix, so the matrix product
$f_1 \, f_2$ should be decomposable as above. Utilizing the
identities Eq.(\ref{ids}), we readily find that the product is
given by
\beqa f_1f_2 = \frac{1}{L^2} \sum_{\bf n} \Delta({\bf n})\otimes
\left( \widehat{f}_1({\bf n}) \star \widehat{f}_2({\bf n})
\right). \nonumber \eeqa
In the right-hand-side, product between the two lattice functions
$\widehat{f_1}({\bf n})$ and $\widehat{f_2}({\bf n})$ defines (the
lattice version of) Moyal's $\star$-product, whose explicit
expression is given by
\beqa \widehat{f}_1({\bf n}) \star \widehat{f}_2({\bf n}) :=
\left[ \widehat{f_1}({\bf n}) \widehat{f_2}({\bf n}) \right]_\star
= \frac{1}{L^2} \sum_{{\bf n}', {\bf n}''}\hspace{-1mm}'\,
\widehat{f}_1({\bf n}') \widehat{f}_2({\bf n}'') \,
e^{2i\widehat{B} ({\bf n}-{\bf n}') \wedge ({\bf n}-{\bf n}'')} \
. \label{moyal1} \eeqa
The parameter $\widehat{B}$ denotes a counterpart of the
`background Neveu-Schwarz B-field' of Type II string theories in
the Seiberg-Witten limit \cite{seibergwitten}:
\beqa \widehat{B} := \left(\Theta' \frac{L^2}{2 \pi} \right)^{-1}
\ . \label{B} \eeqa
The summations over ${\bf n}'$ and ${\bf n}''$ are restricted to
the lattice points satisfying either
$\frac{2}{L\Theta'}(n_a - n'_a)\in {\bf Z}$
or $\frac{2}{L\Theta'}(n_a - n''_a)\in {\bf Z}$.

Rewriting the mother theory action in terms of the hatted
configuration space fields, we finally arrive at the following
form of the noncommutative daughter theory action:
 \beqa
L_g & = & \frac{1}{16g^2}{n \over m} \sum_{\bf n} \tr
\left.\left[\widehat{W}^{\alpha}({\bf n})\widehat{W}_{\alpha}({\bf
n}) \right]_\star \right|_{\theta\theta} + \mbox{h.c.}, \n
L_{\Phi} & = & \frac{1}{g^2}\frac{n}{m} \sum_{\bf n} \tr \left[
\overline{\widehat{\Phi}}_1({\bf n}) e^{\widehat{\cal V}({\bf n})}
\widehat{\Phi}_1({\bf n}) e^{-\widehat{\cal V}({\bf n} +
2\widehat{x})} \right. + \overline{\widehat{\Phi}}_2({\bf n})
 e^{\widehat{\cal V}({\bf n})}
\widehat{\Phi}_2({\bf n}) e^{-\widehat{\cal V}({\bf n}
-\widehat{x}+\widehat{y})} \n
 & & \hspace{1.8cm} \left.\left.
+\overline{\widehat{\Phi}}_3({\bf n}) e^{\widehat{\cal V}({\bf
n})} \widehat{\Phi}_3({\bf n}) e^{-\widehat{\cal V}({\bf n}
-\widehat{x}- \widehat{y})} \right]_\star
\right|_{\theta\theta\bar{\theta}\bar{\theta}}, \n
L_W & = & \frac{\sqrt{2}}{g^2}\frac{n}{m}\sum_{\bf n} \tr \Big[
\omega_I^{-1}\widehat{\Phi}_1({\bf n})\widehat{\Phi}_2({\bf n} +
2\widehat{x}) \widehat{\Phi}_3({\bf n} +\widehat{x} +\widehat{y})
\n
 & & \hspace{1.9cm} \left. -\omega_I \,
\widehat{\Phi}_1({\bf n})\widehat{\Phi}_3({\bf n} + 2\widehat{x})
\widehat{\Phi}_2({\bf n} +\widehat{x} -\widehat{y}) \Big]_\star
\right|_{\theta\theta} + \mbox{h.c.} \ . \label{NC_daughter1}
\eeqa
The daughter theory Eq.(\ref{NC_daughter1}) is a counterpart of
the daughter theory Eq.(\ref{daughter1}). Here, the subscript
$\star$'s refer to (lattice version of) Moyal's $\star$-product
among all the fields inside the square-bracket. In $L_W$, due to
the noncommutativity between $D_1$ and $D_2$, nontrivial
phase-factors $\omega_I^{-1}$ and $\omega_I$ emerged. These
phase-factors correspond to a U(1) magnetic flux penetrating
through each triangular plaquette defined by each of the two
super-potential terms in $L_W$, and are reminiscent of the
discrete torsion in string theory \cite{discretetorsion} (see also
\cite{fabinger}).

As in the commutative case, we shall take $L$ to be odd, and
re-scale the lattice spacings so that (discrete subgroup of) the
two-dimensional rotation symmetry is better represented:
$\epsilon_x=\frac12\epsilon$ and
$\epsilon_y=\frac{\sqrt{3}}{2}\epsilon$. Denote coordinates of the
resulting equilateral triangular lattice as ${\bf x} = (x, y) =
(n_x \epsilon_x, n_x \epsilon_y)$, where $n_x, n_y$ range over the
values specified in Eqs.(\ref{n1}, \ref{n2}). Denote also $\ell
\equiv L \epsilon, \, \ell_x \equiv L \epsilon_x, \, \ell_y \equiv
L \epsilon_y$. The dual basis $\Delta({\bf n})$ is then
transcribed in the continuum limit into
\beqa \Delta({\bf x}) = \sum_{\bf m}  J({\bf m}) \,
{\omega_{\ell_x}}^{m_1 x} {\omega_{\ell_y}}^{m_2 y} \qquad {\rm
where} \qquad \omega_{\ell_{x}} := e^{ 2 \pi i \over \ell_{x}} \,
\quad \omega_{\ell_{y}} := e^{ 2 \pi i \over \ell_{y}} \nonumber
\eeqa
obeying periodicity:
\beqa \Delta({\bf x} + \ell\widehat{x}) =
\Delta\left({\bf x} + \ell_x \widehat{x} + \ell_y
\widehat{y}\right) = \Delta({\bf x}) \ . \nonumber \eeqa
Similarly, the algebraic identities Eq.(\ref{ids}) are transcribed
into:
\beqa
 & & \frac{1}{L^2}\sum_{\bf x} \Delta({\bf x}) = \id_{mq\cdot nq},
 \n
 & & D_1\left[\Delta({\bf x})\otimes \varphi({\bf x})\right]
D_1^{\dagger} = \Delta\left({\bf
x}-\widehat{x}\epsilon_x\right)\otimes\varphi({\bf x}), \n
 & & D_2 \left[\Delta({\bf x})\otimes \varphi({\bf x})\right]
 D_2^{\dagger} =
\Delta\left({\bf x} - \widehat{y}
\epsilon_y\right)\otimes\varphi({\bf x}), \n
 & & \widehat{\tr} \left[ \Delta({\bf x})\right] = mq \cdot nq, \n
 & & \widehat{\tr} \left[ \Delta({\bf x})\Delta({\bf x}') \right] =
  mq \cdot nq \, L^2 \, \delta_{{\bf x}, {\bf x}'}\ , \nonumber \eeqa
for an arbitrary $(k \times k)$ matrix-valued field $\varphi({\bf
x})$, and the (shifted) matrix variable decomposition transcribed
into fields living on two-dimensional noncommutative space:
\beqa \widetilde{\Phi}_s = \frac{1}{L^2} \sum_{\bf x}\Delta({\bf
x})\otimes {\Phi}_s({\bf x}). \nonumber \eeqa
Re-expressing the daughter theory in terms of these fields, the
daughter theory action becomes
\beqa L_g & = & \frac{1}{16g^2}\frac{n}{m} \sum_{\bf x} \tr \left.
\Big[W^{\alpha}({\bf x})W_{\alpha}({\bf x}) \Big]_{\star}
\right|_{\theta\theta} + \mbox{h.c.}, \n L_{\Phi} & = &
\frac{1}{g^2} \frac{n}{m} \sum_{\bf x} \,\, \tr \left[
\overline{\Phi}_1({\bf x}) e^{{\cal V}({\bf x})} \Phi_1({\bf x})
e^{-{\cal V}({\bf x} + {\bf m}_1\epsilon)} \right. +
\overline{\Phi}_2({\bf x}) e^{{\cal V}({\bf x})} \Phi_2({\bf x})
e^{-{\cal V}({\bf x} + {\bf m}_2\epsilon)} \n
 & & \hspace{1.9cm} \left.\left.
+\overline{\Phi}_3({\bf x}) e^{{\cal V}({\bf x})} \Phi_3({\bf x})
e^{-{\cal V}({\bf x} +{\bf m}_3\epsilon)}\right]_{\star}
\right|_{\theta\theta\bar{\theta}\bar{\theta}}, \n L_W & = &
\frac{\sqrt{2}}{g^2}\frac{n}{m} \sum_{\bf x} \tr
\Big[\omega_I^{-1} \,\Phi_1({\bf x})\Phi_2({\bf x} + {\bf
m}_1\epsilon) \Phi_3({\bf x} -{\bf m}_3\epsilon) \n
 & & \hspace{2.0cm} \left. -\omega_I \,
\Phi_1({\bf x})\Phi_3({\bf x} + {\bf m}_1\epsilon) \Phi_2({\bf x}
-{\bf m}_2\epsilon) \Big]_{\star} \right|_{\theta\theta} +
\mbox{h.c.}. \label{NC_daughter2} \eeqa
The noncommutative daughter theory Eq.(\ref{NC_daughter2}) is a
direct counterpart of the commutative daughter theory
Eq.(\ref{daughter2}). Let us contrast salient features of the
latter theory when viewed as a deformation of the former theory.
First, product among fields are deformed into Moyal's
$\star$-product. It is readily found that, in terms of continuum
variables, Moyal $\star$-product is given by
\beqa \widehat{f}_1({\bf x}) \star \widehat{f}_2({\bf x}) &:=&
\frac{1}{mq \cdot nq} \, \widehat{\tr}  \Big(f_1f_2\Delta({\bf
x})\Big) \n &=& \frac{1}{L^2} \sum_{{\bf x}', {\bf
x}''}\hspace{-1mm}' \, f_1({\bf x}') f_2({\bf x}'') \, e^{2iB
({\bf x}-{\bf x}') \wedge ({\bf x}-{\bf x}'')} \ ,  \label{moyal2}
\eeqa
where
\beqa
 B := \left( \Theta' {\ell_x \ell_y \over 2 \pi} \right)^{-1}
 \nonumber
 \eeqa
and the sums over ${\bf x}'$ and ${\bf x}''$ are restricted to
either
$$
\left\{\frac{2}{\ell_x\Theta'}(x - x')\in {\mathbf Z}, \,
\frac{2}{\ell_y\Theta'}(y - y')\in {\bf Z}\right\} \mbox{ or }
\left\{\frac{2}{\ell_x\Theta'}(x - x'')\in {\mathbf Z}, \,
\frac{2}{\ell_y \Theta'}(y - y'')\in {\bf Z}\right\}.
$$
We emphasize that the noncommutativity, as defined through Moyal's
product, is determined not by $\Theta$ but by $\Theta'$. Second,
the gauge coupling parameter is re-scaled as
\beqa g^2 \quad \longrightarrow \quad g^2_{\rm NC} := g^2 \left({m
\over n} \right) \ . \nonumber \eeqa
Third, as the noncommutative deformation is made, nontrivial
phase-factors $\omega_I^{-1}$ and $\omega_I$, respectively, are
induced on the two terms in the superpotential $L_W$. One might
consider these phase-factors trivial as they seem to disappear in
the limit $m, n \rightarrow \infty$. In the next subsection, we
will find that they actually retain nontrivial effects in the
continuum limit.

Note that the nontrivial phase-factors do not affect the
R-symmetry of the noncommutative daughter theory {\sl on the
lattice}. It is $G_R^{\rm NC \,\, daughter} = [{\rm U}(1)]^3
\times {\rm Spin}(3)$, same as the one in the commutative
counterpart.
Again, the diagonal U(1) corresponds to the R-symmetry
of the manifest supersymmetry on the two-dimensional
noncommutative lattice, as should be the same as that of
(2+1)-dimensional ${\cal N}=2$ supersymmetry.
\subsection{(Classical) Continuum Limit}
We now take a continuum limit of the Lagrangian
Eq.(\ref{NC_daughter2}):
\beqa \epsilon \rightarrow 0, \quad \ell_{x,y} = L \epsilon_{x,y}
= \mbox{fixed}, \quad m, \, n \rightarrow \infty, \qquad {m \over
n} := \mu = \mbox{fixed}. \label{contlimit}\eeqa
The last condition is to ensure the coupling parameter $g^2_{\rm
NC}$ and the noncommutativity finite. As we will see shortly,
non-trivial phase factors survive in the limit.

We split the superpotential $L_W$ in Eq.({\ref{NC_daughter2}) into
the two parts $L_W = L_W^{(c)} + L_W^{(s)}$ (not real and
imaginary parts), where
\beqa L_W^{(c)} & = & \frac{\sqrt{2}}{g^2_{\rm NC}}
\left(\cos\frac{2\pi}{I}\right) \sum_{\bf x} \tr \Big[\Phi_1({\bf
x})\Phi_2({\bf x} + {\bf m}_1\epsilon) \Phi_3({\bf x} -{\bf
m}_3\epsilon) \n
 & & \hspace{3.2cm} - \left.
\Phi_1({\bf x})\Phi_3({\bf x} + {\bf m}_1\epsilon) \Phi_2({\bf x}
-{\bf m}_2\epsilon)\Big]_{\star} \right|_{\theta\theta} \n
 & & + \mbox{h.c.}, \label{costerm} \\
L_W^{(s)} & = & \frac{\sqrt{2}}{g^2_{\rm NC}}
\left(-i\sin\frac{2\pi}{I}\right) \sum_{\bf x} \tr
\Big[\Phi_1({\bf x})\Phi_2({\bf x} + {\bf m}_1\epsilon)
\Phi_3({\bf x} -{\bf m}_3\epsilon) \n
 & & \hspace{3.2cm} + \left.
\Phi_1({\bf x})\Phi_3({\bf x} + {\bf m}_1\epsilon) \Phi_2({\bf x}
-{\bf m}_2\epsilon)\Big]_{\star}\right|_{\theta\theta} \n
 & & + \mbox{h.c.} . \label{sinterm}
\eeqa
In the continuum limit, structure of $L_g$, $L_{\Phi}$,
$L_W^{(c)}$ coincide with $L_g$ , $L_{\Phi}$, $L_W$ in
Eq.(\ref{continuum}), except that all the product are replaced by
the (continuum version of) Moyal's $\star$-product:
\beqa f_1({\bf x})\star f_2({\bf x}) = \left. e^{\frac{i}{2}
\widehat{\Theta}'
(\partial_{x_1}\partial_{y_2}-\partial_{x_2}\partial_{y_1})}f_1({\bf
x})f_2({\bf y}) \right|_{y \rightarrow x} \ . \nonumber \eeqa
A continuum version of the non-commutativity parameter is given by
$\widehat{\Theta}'$:
\beqa \widehat{\Theta}' \equiv \Theta' {\ell_x \ell_y \over 2 \pi}
= {1 \over B}. \nonumber \eeqa
We may also take an infinite volume limit while holding the
non-commutativity $\widehat{\Theta}'$ finite
\cite{mandalreywadia}:
\beqa \ell_x \ell_y \rightarrow \infty, \qquad q \rightarrow
\infty, \qquad r \ll \mu \ . \label{infvolume} \eeqa

The term $L_W^{(s)}$ proportional to $\sin \frac{2\pi}{I}$, on the
other hand, gives rise to a nontrivial term in the continuum
limit, and, in the limit Eq.(\ref{infvolume}), yields
\beqa L_W^{(s)} &=& -i \frac{1}{g_{\rm NC}^2}\left(\frac{2\pi
q}{\ell_x\ell_y}\frac{m}{n}\right) \int {\rm d}^2
x~\frac{1}{\sqrt{6}}~\tr(F_1+F_2+F_3) + \mbox{h.c.} \n &:=& \left.
-i \int {\rm d}^2 x \, {\rm tr} \Big[\xi_1 {\Phi}_1 + \xi_2
{\Phi}_2 + \xi_3 {\Phi}_3 \Big] \right|_{\theta \theta} +
\mbox{h.c.} \label{phase} \eeqa
with
\beqa \xi_1 = \xi_2 = \xi_3 \sim {1 \over \widehat{\Theta}'} = B.
\nonumber \eeqa
As such, Eq.(\ref{phase}) represents turning on in the
super-potential terms linear in the scalar fields arising from the
link variables. Combining with the result for $L_W^{(c)}$, we see
that Eq.(\ref{phase}) represents a magnetic flux background
$F_{xy} = - B \id_k$ of the diagonal U(1) gauge group.

One marked difference of the noncommutative daughter theory is
that, unlike the commutative counterpart, the lattice
supersymmetry is not enhanced to the full sixteen supercharges in
the continuum limit, but remains same as that at finite lattice
spacing. This may be seen, for example, from symmetry mismatch
between Eq.(\ref{costerm}) and Eq.(\ref{phase}). The cubic
superpotential terms in Eq.(\ref{costerm}) are invariant in the
continuum limit under enhanced R-symmetry group
Spin(4)$\times$SO(2), which is a subgroup of Spin(6) transforming
real and imaginary parts of ${\Phi}_j$ $(j=1,2,3)$ in the defining
representation. The individual phase rotations $[{\rm U}(1)]^3$ of
$\Phi_j$'s belong to the R-symmetry group.
On the other hand, the linear superpotential
terms in Eq.(\ref{phase}), which descend from Eq.(\ref{sinterm})
in the continuum limit, are invariant under the ${\rm U}(1)$ group,
rotating phases of $\Phi_j$'s simultaneously:
\beq
\Phi_j \rightarrow e^{i2\beta}\Phi_j \qquad {\rm with} \qquad
\theta \rightarrow e^{-i\beta}\theta.
\label{diagonalU(1)2}
\eeq
This rotation (\ref{diagonalU(1)2}),
however, does not belong to (any linear combination of)
the $[{\rm U}(1)]^3$. As such, the R-symmetry of the continuum
theory, consisting of linear and cubic terms in the
superpotential, is not present,
except Spin(3) keeping $\Phi_j$'s intact.
It indicates at most four conserved
supercharges in the continuum limit.

The definition of Moyal's $\star$-product Eq.(\ref{moyal2})
indicates that, in case $\Theta'$ is an even-integer, the
noncommutative daughter theory reduces to the ordinary commutative
one. However, the linear superpotential terms $L_W^{(s)}$
in Eq.(\ref{phase}) remain non-vanishing, and hence do not have
enhancement of the supersymmetry. It just means that the limit
$\widehat{\Theta}'$ gives rise to a different commutative theory
(with four supercharges only) from the one in
Eq.(\ref{continuum}).

\subsection{Twist the Mother Theory!}
Consideration of the (classical) continuum limit given above
suggests the following possibility. Suppose we start with a
`twisted' version of the mother theory Eq.(\ref{mother}), where
the only difference from Eq.(\ref{mother}) would be that terms in
the superpotential $L_W$ are modified by a phase-factor $z$:
\beqa L^{\rm twisted} & = & L_g + L_{\Phi} + L_W^{\rm twisted} ,
\n L_W^{\rm twisted} & = & \left. \frac{\sqrt{2}}{g^2} \, \Tr \,
\Big[z\Phi_1\Phi_2\Phi_3 - z^{-1}\Phi_1\Phi_3\Phi_2 \Big]
\right|_{\theta\theta} + \mbox{h.c.}. \nonumber \eeqa
Note that, because of the phase-factors $z, z^{-1}$ introduced,
R-symmetry of the `twisted' mother theory with $z\neq 1$ differs
from the untwisted one --- the untwisted mothery theory possesses
sixteen supercharges, while the twisted mother theory would
possess only four. For both theories, after the orbifold
conditions Eq.(\ref{NC_orbifolding}) are imposed, the same four
supercharges are retained. Repeating the same analysis as in the
previous section, we obtain the following twisted, noncommutative
daughter theory:
\beqa L_W^{\rm twisted} & = & \frac{\sqrt{2}}{g^2}\frac{n}{m}
\sum_{\bf x} \tr \Big[ z\omega_I^{-1} \Phi_1({\bf x})\Phi_2({\bf
x} + {\bf m}_1\epsilon) \Phi_3({\bf x} -{\bf m}_3\epsilon) \n
 & & \hspace{1.6cm} \left. -z^{-1}\omega_I
\Phi_1({\bf x})\Phi_3({\bf x} + {\bf m}_1\epsilon) \Phi_2({\bf x}
-{\bf m}_2\epsilon) \Big]_{\star} \right|_{\theta\theta} +
\mbox{h.c.}, \nonumber \eeqa
while $L_g$ and $L_{\Phi}$ remains the same as in the untwisted
ones in Eq.(\ref{NC_daughter2}). For a generic choice of the
phase-factor $z$, the four supercharges manifest on the lattice is
not enhanced in the continuum limit. Interestingly, at the special
choice of $z=\omega_I$, however, the lattice supersymmetry gets
enhanced to preserve the full sixteen supercharges in the
continuum limit, as the extra superpotential term Eq.(\ref{phase})
disappears. The R-symmetry Spin(7) emerges in the continuum limit
accordingly, much the same as in the commutative case. Intuitively
speaking, the choice $z = \omega_I$ amounts to turning on
background gauge field flux along the diagonal U(1) subgroup in
the mother theory so that it counter-balances out gauge flux of
the daughter theory induced via the linear superpotential terms
Eq.(\ref{phase}).

\section{Discussions}
In this paper, we have studied gauge theories with manifest
supersymmetry on a noncommutative lattice. We have provided a
systematic prescription for constructing noncommutative
space-time, and have constructed explicitly a $(2+1)$-dimensional
noncommutative lattice gauge theory as a daughter theory of
$(0+1)$-dimensional mother gauge theory. Extensions to
$(d+1)$-dimensional mother theory and to lattices of more than
two dimensions are straightforward.

A notable feature of constructing Yang-Mills theories out of
matrices via orbifold condition is that rank of the gauge group of
the Yang-Mills theories can be taken finite (though the size of
the matrices should be necessarily taken to infinity in order to
ensure continuum and infinite volume limits). This would be a
significant advantage over the traditional Eguchi-Kawai reduction
\cite{EK}, which is limited only to the planar limit of the
Yang-Mills theory.

Another novel feature not encountered in the commutative
counterpart (section 3) is that the total number of preserved
supercharges in the continuum limit depends on a deformation
parameter one can introduce to the mother theory. The supercharges
in the continuum limit is generically four, the same amount of the
supersymmetry as the lattice theory, but is enhanced to sixteen at
a particular value of the deformation parameter. As anticipated,
the parameter is determined solely by the noncommutativity.

There are several issues deserving further study. One is
concerning nonperturbative definition of physical observables. In
the continuum formulation and in the super-gravity dual
formulation via AdS/CFT correspondence, it was shown
\cite{kawaietal, dasunge, dasrey, grossetal, Rey:nq} that, due to
novelty of the noncommutative gauge invariance, physical
observables are `open Wilson lines', operators local in momentum
space but nonlocal in configuration space. It would be very
interesting to understand how these open Wilson lines emerge out
of the orbifold conditions. Another is regarding potential
relation between the noncommutativity and the discrete torsion
that has shown up prominently in string theories. Identification
of a precise relation would help understanding the discrete
torsion intuitively. Finally, results in this work are based on a
naive continuum limit dealing with the action itself. To
understand the limit in full detail, one would need to understand
nontrivial issues of proper renormalization of operators, at least
in the lattice perturbation theory \cite{Nishimura:2002hw}, in
noncommutative gauge theories. Given that our results are rigorous
at least in the case that $\Theta'$ is even-integer, where the
theory reduces to a commutative one (with background gauge flux),
one would expect that these issues are amenable with little
technical difficulty.


\renewcommand{\theequation}{\Alph{section}.\arabic{equation}}
\appendix
\section{Superfield Notation}
\setcounter{equation}{0}
 In the Wess-Zumino gauge, we denote the (3+1)-dimensional ${\cal N} = 1$
 superfield (the vector superfield ${\cal V}$
and the chiral superfield $\Phi$) in terms of the component
fields.

\beqa {\cal V}(x) & = & -2\theta\sigma^m\bar{\theta}v_m(x)
+2i\theta\theta\bar{\theta}\bar{\lambda}(x)
-2i\bar{\theta}\bar{\theta}\theta\lambda(x)
+\theta\theta\bar{\theta}\bar{\theta}D(x), \n \Phi(x) & = & B(y) +
\sqrt{2}\theta\psi(y) + \theta\theta F(y), \nonumber \eeqa
where $y^m = x^m +i\theta\sigma^m\bar{\theta}$, $m=0,\cdots, 3$.
$v_m$ are gauge fields and $\lambda$, $\bar{\lambda}$ are gaugino.
$B$ is a complex scalar field, and $\psi$ is its super-partner.
$D$ and $F$ are auxiliary fields.

$W_{\alpha}$ is a chiral superfield containing the field strength;
\beqa W_{\alpha}(x) = -2i\lambda_{\alpha}(y) +
2\theta_{\alpha}D(y) -2i(\sigma^{mn}\theta)_{\alpha}v_{mn}(y) +
2\theta\theta(\sigma^m{\cal D}_m\bar{\lambda}(y))_{\alpha},
\nonumber \eeqa
where the spinor-index $\alpha$ runs over 1,2. Also
\beqa v_{mn} & = & \partial_mv_n - \partial_nv_m + i[v_m, v_n], \n
{\cal D}_m\bar{\lambda} & = & \partial_m\bar{\lambda} +i[v_m,
\bar{\lambda}]. \nonumber \eeqa
%

\acknowledgments
We thank Sergei Cherkis, Tohru Eguchi, Hikaru Kawai, Kyungkyu Kim,
Sooil Lim, Herbert Neuberger, Nathan Seiberg and Shigeki Sugimoto
for enlightening discussions. J.N. thanks the String Theory Group
at Seoul National University for warm hospitality, where this work
was initiated. The work of J.N.\ is supported in part by
Grant-in-Aid for Scientific Research (No.\ 14740163) from the
Ministry of Education, Culture, Sports, Science and Technology.
The work of S-J.R. and F.S. is supported in part by the KOSEF
Interdisciplinary Research Grant 98-07-02-07-01-5, the KOSEF
Leading Scientist Grant, and the Fellowship from the Institute for
Advanced Study.

\end{document}